\documentclass{amsart}
\usepackage{graphicx}
\usepackage{amssymb}
\usepackage{amsfonts}
\swapnumbers
\newtheorem{theorem}{Theorem}[section]

\theoremstyle{definition}
\newtheorem{definition}[theorem]{Definition}

\numberwithin{equation}{section}
 \theoremstyle{plain}
 
 \numberwithin{equation}{section} 
 \numberwithin{figure}{section} 
 \theoremstyle{plain}
 \theoremstyle{plain}
 \theoremstyle{remark}
 \newtheorem*{acknowledgement*}{Acknowledgement}

\newcommand{\cA}{{\mathcal A}}

\newcommand{\cF}{{\mathcal F}}

\newcommand{\cP}{{\mathcal P}}

\newcommand{\cT}{{\mathcal T}}

\newcommand{\Om}{{\Omega}}

\newcommand{\Del}{{\Delta}}
\newcommand{\gam}{{\gamma}}
\newcommand{\Gam}{{\Gamma}}

\newcommand{\sig}{{\sigma}}
\newcommand{\al}{{\alpha}}
\newcommand{\be}{{\beta}}
\newcommand{\ka}{{\kappa}}
\newcommand{\la}{{\lambda}}

\newcommand{\bbR}{{\mathbb R}}

\newcommand{\bbI}{{\mathbb I}}

\begin{document}
\title[]{Shortfall Minimization for Game Options in Discrete Time}%
 \vskip 0.1cm
 \author{ Yuri Kifer\\
 Institute of Mathematics, Hebrew University of Jerusalem}%
\address{}%

\thanks{ }
\dedicatory{  }
 \date{\today}
\begin{abstract}\noindent
We prove existence of a self-financing strategies which minimize shortfall
for game options in discrete time.
\end{abstract}
\maketitle
\markboth{Yu.Kifer}{Shortfall Minimization}
\renewcommand{\theequation}{\arabic{section}.\arabic{equation}}
\pagenumbering{arabic}

\section{Definitions and results}\setcounter{equation}{0}
We consider here Israeli (game) contingent claims introduced in \cite{Ki}.
\begin{definition} Let $\ell$ be a convex increasing function equal zero on $\{ x\leq 0\}$,
$N$ be a positive integer and
 $H(m,n)=U_m\bbI_{m<n}+W_n\bbI_{n\leq m}$, $U_k\geq W_k\geq 0\,\,\,\forall k\geq 0$ be a payoff
 function satisfying $E\ell( U_n)<\infty$ for $n=0,1,...,N$.
 The shortfall risk $r(\pi)$ for a game contingent claim given a self-financing strategy
 $\pi$ is defined by
 \[
 r(\pi)=\inf_{\sig\in\cT_{0N}}\sup_{\tau\in\cT_{0N}}E\ell(H(\sig,\tau)-X^\pi_{\sig\wedge\tau})=
 \inf_{\sig\in\cT_{0N}}\sup_{\tau\in\cT_{0N}}E\ell((H(\sig,\tau)-X_{\sig\wedge\tau}^\pi)^+)
 \]
 where $\cT_{0N}$ is the set of all stopping times $\tau$ taking values between $0$ and $N$.
 \end{definition}
 In the game contingent claim case a self-financing portfolio strategy is called admissible
 if the corresponding portfolio value $X_n^\pi$ is nonnegative for all $n=0,1,...,N$. The set of
 admissible strategies with $X_0^\pi\leq x$ will be denoted by $\Pi_x$.  The existence of an
 admissible portfolio strategy $\pi^*$ such that $r(\pi^*)=inf_{\pi\in\Pi_x}r(\pi)$ for
 $x>0,\, x<\inf_{\sig\in\cT_{0N}}\sup_{\tilde P\in\cP(P),\,\tau\in\cT_{0N}}E_{\tilde P}(H(\sig,\tau))$
  for markets defined on a finite probability space is not difficult to obtain. Indeed,
  for any $\sig\in\cT_{0N}$ and $\pi\in\Pi_x$ set
  \[
  r(\sig,\pi)=\sup_{\tau\in\cT_{0N}}E\ell(H(\sig,\tau)-X^\pi_{\sig\wedge\tau}).
  \]
  Then, in the same way as in the American contingent claims case we can find $\pi_\sig\in\Pi_x$
  such that
  \[
  r(\sig,\pi_\sig)=\inf_{\pi\in\Pi_x}r(\sig,\pi).
  \]
  Taking into account that there exist only finitely many stopping times in $\cT_{0N}$ when the sample
  space $\Om$ is finite (or when the filtration consists of finite $\sig$-algebras) we obtain that there
  exists $\pi^*\in\Pi_x$ such that
  \[
  r(\pi^*)=\min_{\sig\in\cT_{0N}}r(\sig,\pi_\sig)=\inf_{\pi\in\Pi_x}r(\pi).
  \]
  For general markets the existence of such $\pi^*$ is a more difficult question due, in particular, to the fact
   that $r(\pi)$ defined above is not a convex function of $\pi$ in this case. Still, it turns out that
   the question can be answered affirmatively also for game contingent claims which will extend the results
   of Section 6 in \cite{DK0}.

We consider a discrete time market on a general probability space $(\Om,\cF,P)$ with discounted quantities
so that the bond price $B_n\equiv 1,\,\forall n\geq 0$ stays constant while the stock price at time $n$
has the form
\[
S_n=S_0\prod_{k=1}^n(1+\rho_k)
\]
where $S_0$ is a positive constant and $\rho_1,\rho_2,...$ is a sequence of random variables with values
in $(-1,\infty)$. We assume also that the market admits no arbitrage.

Next, consider a game contingent claim with the horizon $N<\infty$ and the payoff function
\[
H(m,n)=U_m\bbI_{m<n}+W_n\bbI_{n\leq m}
\]
where $U_k=f_k(\rho_1,...,\rho_k)\geq W_k=g_k(\rho_1,...,\rho_k)\geq 0$ for some measurable functions $f_k=f_k(x_1,...,x_k)\geq g_k=g_k(x_1,...,x_k)$ on $\bbR^k$, $k=1,...,N$.
 We assume that $E\ell(U_k)<\infty$ for all $k=1,2,...N$.

\begin{theorem}\label{shortfall}
Under the above conditions for any $x>0$ there exists a stopping time $\sig^*\leq N$ and an admissible
 self-financing portfolio strategy $\pi^*$ such that
 \[
 r(\sig^*,\pi^*)=\inf_{\sig\in\cT_{0N},\pi\in\Pi_x}r(\sig,\pi).
 \]
 \end{theorem}

 \section{Proofs}\setcounter{equation}{0}
 \begin{proof} Set $\rho^{(k)}=(\rho_1,...,\rho_k)$ and denote by $\mu^{(k)}$ the distribution of the
 random vector $\rho^{(k)}$ on $(-1,\infty)^k\subset\bbR^k$. Let $\cF_0=\{\emptyset,\Om\}$,
 $\cF_k=\sig\{\rho_1,...,\rho_k\}$, $k=1,2,...$ be $\sig$-algebras generated by $\rho_1,\rho_2,...$. It is
 well known (see, for instance, Theorems 10.2.1 and 10.2.2 in \cite{Du}) that in this circumstances for
 each $x^{(k)}=(x_1,...,x_k)\in(-1,\infty)^k$ there exists a probability measure $\mu_{x^{(k)}}^{(k+1)}$
 on $(-1,\infty)$ such that for any Borel set $\Gam\subset\bbR$,
 \[
 \mu_{x^{(k)}}^{(k+1)}(\Gam)=P\{\rho_{k+1}\in\Gam |\cF_k\}\quad P-\mbox{a.s.}
 \]
 and for any Borel set $Q\subset\bbR^{k+1}$,
 \[
 \mu^{(k+1)}(Q)=\int_{\bbR^k}\mu^{(k+1)}_{x^{(k)}}(Q_{x^{(k)}})d\mu^{(k)}(x^{(k)})
 \]
 where $Q_{x^{(k)}}=\{ x\in\bbR:\, (x^{(k)},x)\in Q\}$. Such measures $\mu_{x^{(k)}}^{(k+1)}$ are called
 regular conditional probabilities or disintegrations of the measure $\mu^{(k+1)}$. Observe that if $\rho_1,
 \rho_2,...$ are independent then $\mu_{x^{(k)}}^{(k+1)}$ does not depend on $x^{(k)}$ and it is equal to
 the distribution $\nu_{k+1}$ of $\rho_{k+1}$, so that in this case $\mu^{(k+1)}=\nu_1\times\nu_2\times\cdots
 \times\nu_{k+1}$.

Recall, that the discounted portfolio value at time $n$ corresponding to a self-financing portfolio
 strategy $\pi=(\be_n,\gam_n),\, n\geq 0$ can be written in the form
 \[
 X_n^\pi=\be_n+\gam_nS_n=X_0^\pi+\sum_{k=1}^n\gam_k\Del S_k.
 \]
 Since the market admits no arbitrage then,
 \begin{equation*}
 \mu_{x^{(n)}}^{(n+1)}(0,\infty)<1\quad\mbox{for $\mu^{(n)}$-almost all}\,\,\, x^{(n)}
 \end{equation*}
 since for otherwise we could make a riskless profit on $(n+1)$-stage by buying stocks on the $n$-th stage
 when $\rho^{(n)}$ equals $x^{(n)}$ satisfying $\mu_{x^{(n)}}^{(n+1)}(0,\infty)=1$ and selling them on the
 next stage. Thus for $\mu^{(n)}$-almost all $x^{(n)}$ either
  \begin{equation}\label{L10.1}
 \mbox{supp}\mu_{x^{(n)}}^{(n+1)}\cap(-\infty,0)\ne 0\,\,\,\mbox{or}\,\,\,\mbox{supp}\mu_{x^{(n)}}^{(n+1)}=
 \{ 0\}
 \end{equation}
 where supp denotes the support of a measure, i.e. the complement of the union of all open sets of
 zero measure. Let
 \[
 a_{n+1}(x^{(n)})=\inf(\mbox{supp}\mu_{x^{(n)}}^{(n+1)}).
 \]
  If $\pi$ is admissible then with probability one
  \begin{equation}\label{L10.2}
  X^\pi_{n+1}=X_n^\pi+\gam_{n+1}\Del S_{n+1}=X_n^\pi+\gam_{n+1}S_n\rho_{n+1}\geq 0,
  \end{equation}
  and so
  \begin{equation}\label{L10.3}
  X^\pi_n+\gam_{n+1}S_na_{n+1}(\rho^{(n)})\geq 0.
  \end{equation}
  By (\ref{L10.1}) and (\ref{L10.3}) with probability one either $-\infty\leq a_{n+1}(\rho^{(n)})<0$
  and
  \[
  \gam_{n+1}\leq\frac {X^\pi_n}{S_n(-a_{n+1}(\rho^{(n)}))}\,\,\mbox{where}\,\,\frac C{-\infty}=0,
  \]
  or $a_{n+1}(\rho^{(n)})=0$ and then $\mbox{supp}\mu_{\rho^{(n)}}^{(n+1)}=\{ 0\}$ and
  any choice of $\gam_{n+1}\geq 0$ will preserve admissibility.

 Next, set
 \[
 b_{n+1}(x^{(n)})=\sup(\mbox{supp}\mu_{x^{(n)}}^{(n+1)}).
 \]
 and observe that for $\mu^{(n)}$-almost all $x^{(n)}$ either
  \[
  \mbox{supp}\mu^{(n+1)}_{x^{(n)}}\cap(0,\infty)\ne 0\,\,\,\mbox{or}\,\,\,\mbox{supp}\mu^{(n+1)}_{x^{(n)}}=
  \{ 0\}
  \]
  since for otherwise we could make a riskless profit on the $(n+1)$-th stage by short selling the stocks on
  the $n$-th stage when $\mu^{(n+1)}_{\rho^{(n)}}(-\infty,0)=1$ and closing these positions on the next stage.
  By (\ref{L10.2}),
 \[
  X^\pi_n+\gam_{n+1}S_nb_{n+1}(\rho^{(n)})\geq 0.
  \]
  Hence, with probability one either $0<b_{n+1}(\rho^{(n)})\leq\infty$ and then
  \begin{equation*}
  \gam_{n+1}\geq -\frac {X^\pi_n}{S_nb_{n+1}(\rho^{(n)})}\,\, \mbox{where}\,\, \frac C\infty=0,
  \end{equation*}
  or $b_{n+1}(\rho^{(n)})=0$ and then $\mbox{supp}\mu_{\rho^{(n)}}^{(n+1)}=\{ 0\}$ and any choice
  of $\gam_{n+1}\leq 0$ will preserve admissibility.
  The above conditions on $\gam_{n+1}$ are the only constraints which keep the portfolio value nonnegative
  on the next $(n+1)$-th stage. This discussion motivates to consider the set $\cA_n(X,\rho^{(n)})$ of all
  possible portfolio values at the time $n+1$ provided that the portfolio value at the time $n$ was $X$ which
  is a nonnegative $\cF_n$-measurable random variable. Hence,
  \begin{eqnarray*}
  &\cA_n(X,\rho^{(n)})=\big\{ Y:\, Y=X+\al\Del S_{n+1}\,\,\mbox{for}\,\,  -\frac {X}{S_nb_{n+1}(\rho^{(n)})}\leq \al\leq -\frac {X}{S_na_{n+1}(\rho^{(n)})}\\
  &\mbox{if}\,\, 0> a_{n+1}(\rho^{(n)}\geq -\infty\,\,\mbox{and}\,\,0< b_{n+1}(\rho^{(n)}\leq\infty\\
  &\mbox{while}\,\, Y=X\,\,\mbox{if}\,\, a_{n+1}(\rho^{(n)})=0\,\,\mbox{or}\,\,b_{n+1}(\rho^{(n)})=0\big\}.
  \end{eqnarray*}

Next, we introduce the following optimal stopping (Dynkin's) game. For each admissible portfolio strategy
$\pi\in\Pi_x,\, x>0$ set
\[
Q^\pi(m,n)=\ell(U_m-X^\pi_m)\bbI_{m<n}+\ell(W_n-X_n^\pi)\bbI_{m\geq n}
\]
where, as before, $\ell$ is a convex increasing loss function equal zero on $\{ x\leq 0\}$ and such that
$E\ell(U_n)<\infty$ for all $n\geq 0$. Define by the backward induction $\Psi_N^\pi=\ell(U_N-X_N^\pi)$
and for $n=N-1,N-2,...,0$,
\[
\Psi_n^\pi=\min\big(\ell(U_n-X^\pi_n),\,\max(\ell(W_n-X^\pi_n),E(\Psi^\pi_{n+1}|\cF_n))\big),
\]
where, recall, $U_n=f_n(S_n)$ and $W_n=g_n(S_n)$. Then by the results on Dynkin games discussed in
Lecture 3 we obtain that
\[
\Psi^\pi_0=\inf_{\sig\in\cT_{0N}}\sup_{\tau\in\cT_{0N}}EQ^\pi(\sig,\tau)=r(\pi).
\]
Moreover, there exists $\sig=\sig(\pi)$ such that
\[
r(\pi,\sig)=\sup_{\tau\in\cT_{0N}}EQ^\pi(\sig,\tau)=r(\pi).
\]
In what follows we are going to construct $\pi^*\in\Pi_x$ and $\sig^*\in\cT_{0N}$ such that
\begin{equation}\label{optimal}
r(\pi^*,\sig^*)=\inf_{\pi\in\Pi_x}r(\pi).
\end{equation}

Recall that if $f:\,[0,\infty)\to[0,\infty)$ is a lower semi-continuous function, i.e. $\liminf_{z\to z_0}
f(z)\geq f(z)$ for any $z_0$, then
\[
\mbox{arg}\min_{0\leq z\leq a}f(z)=\min\{ 0\leq\tilde z\leq a:\, f(\tilde z)=\min_{0\leq z\leq a}f(z)\}
\]
is well defined. We will need the following functions defined by the backward induction. First, we put
$I_N(x^{(N)},y,z)=J_N(x^{(N)},y)=\ell(f_N(x^{(N)})-y),\, x^{(N)}\in(-1,\infty),y\geq 0$ and then for any
$n<N$, $z\in(-\infty,\infty)$, $y\ge 0$ and $x^{(n)}=(x_1,...,x_n)\in(-1,\infty)^n$ we set
\begin{eqnarray*}
&I_n(x^{(n)},y,z)=\min\big(\ell(f_n(x^{(n)})-y),\\
&\max(\ell(g_n(x^{(n)})-y),\,\int_{-1}^\infty J_{n+1}
((x^{(n)},u),\, y+zu\ka_n(x^{(n)}))d\mu_{x^{(n)}}^{(n+1)}(u)\big)
\end{eqnarray*}
and
\[
J_n(x^{(n)},y)=\inf_{ z\in G_n(x^{(n)},y)}I_n(x^{(n)},y,z)
\]
where $\ka_n(x^{(n)})=S_0\prod_{k=1}^n(1+x_k)$ and
\begin{eqnarray*}
&G_n(x^{(n)},y)=\{ z\in\bbR:\, -y(\ka_n(x^{(n)})b_{n+1}(x^{(n)}))^{-1}\leq z\leq
-y(\ka_n(x^{(n)})a_{n+1}(x^{(n)}))^{-1}\\
&\mbox{if}\,\, 0>a_{n+1}(x^{(n)})\geq -\infty\,\,\mbox{and}\,\, 0<b_{n+1}(x^{(n)})\leq\infty\\
&\mbox{while $z$ is arbitrary if}\,\, a_{n+1}(x^{(n)})=0\,\,\mbox{or}\,\, b_{n+1}(x^{(n)})=0\}.
\end{eqnarray*}
In order to use the above arg$\min$ notion we will need to show that $I_n(x^{(n)},\cdot,\cdot)$ and $J_n(x^{(n)},\cdot)$ are lower semi-continuous in the arguments denoted by dots.

Since $\ell$ is a continuous function, $I_N(x^{(N)},\cdot,\cdot)=J_N(x^{(N)},\cdot)$ are continuous, and so
they are lower semi-continuous. Suppose that the lower semi-continuity is established for $I_n$ and $J_n$ with $n=N,N-1,...,m+1$ and we prove it for $n=m$. Let $\lim_{k\to\infty}y_k=y$ and $\lim_{k\to\infty}z_k=z$. Since
$J_{m+1}((x^{(m)},u),\cdot,\cdot)$ is lower semi-continuous we obtain by the Fatou lemma that
\begin{eqnarray*}
&F_m(x^{(m)},y,z)=\int_{-1}^\infty J_{m+1}((x^{(m)},u),y+zu\ka_m(x^{(m)}))d\mu_{x^{(m)}}^{(m+1)}(u)\\
&\leq\int_{-1}^\infty\liminf_{k\to\infty}J_{m+1}((x^{(m)},u),y_k+z_ku\ka_m(x^{(m)}))d\mu_{x^{(m)}}^{(m+1)}(u)\\
 &\leq\liminf_{k\to\infty}\int_{-1}^\infty J_{m+1}((x^{(m)},u),y_k+z_ku\ka_m(x^{(m)}))d\mu_{x^{(m)}}^{(m+1)}(u)\\
 &=\liminf_{k\to\infty}F_m(x^{(m)},y_k,z_k),
 \end{eqnarray*}
 and so $F_m(x,y,z)$ is lower semi-continuous in $y$ and $z$. Since $\ell$ is a continuous function we obtain
 that $I_{m}(x^{(m)},y,z)$ is lower semi-continuous in $y$ and $z$, as well.

 Next, let $\lim_{k\to\infty}y_k=y$ be
 such that $\lim_{k\to\infty}J_{m}(x^{(m)},y_k)$ exists. If $a_{m+1}(x^{(m)})\ne 0$ and $b_{m+1}(x^{(m)})\ne 0$
 then since $I_{m}(x^{(m)},y,z)$ is lower semi-continuous in $z$,
 for each $y$ and $x^{(m)}\in(-1,\infty)$ there exists $z_k\in G_m(x^{(m)},y_k)$ such that
 $J_{m}(x^{(m)},y_k)=I_{m}(x^{(m)},y_k,z_k)$. When $a_{m+1}(x^{(m)})\ne 0$ and $b_{m+1}(x^{(m)})\ne 0$ then
 the sequence $\{ z_k\}_{k\geq 1}$ stays in a compact region, and so we can choose a convergent subsequence
 $z_{k_i}\to z$ as $i\to\infty$ with $z\in G(x^{(m)},y)$. Then
 \begin{eqnarray*}
& J_{m}(x^{(m)},y)\leq I_{m}(x^{(m)},y,z)\leq\liminf_{i\to\infty}I_{m}(x^{(m)},y_{k_i},z_{k_i})\\
&=\liminf_{i\to\infty}J_{m}(x^{(m)},y_{k_i})=\lim_{k\to\infty}J_{m}(x^{(m)},y_k).
 \end{eqnarray*}
Since for any sequence $y_k\to y$ we can choose a subsequence $y_{k_i}$ such that
\[
\liminf_{k\to\infty}J_{m}(x^{(m)},y_k)=\lim_{i\to\infty}J_{m}(x^{(m)},y_{k_i})
\]
we obtain that
\[
J_{m}(x^{(m)},y)\leq\liminf_{k\to\infty}J_{m}(x^{(m)},y_k)
\]
completing the induction step when $a_{m+1}(x^{(m)})\ne 0$ and $b_{m+1}(x^{(m)})\ne 0$.

If $a_{m+1}(x^{(m)})=0$ or $b_{m+1}(x^{(m)})=0$ then $\mu_{x^{(m)}}^{(m+1)}\{ 0\}=1$, and so
\[
F_m(x^{(m)},y,z)=J_{m+1}((x^{(m)},0),y)
\]
does not depend on $z$, whence $I_m(x^{(m)},y,z)$ does not depend on $z$ in this case, as well.
Then $J_m(x^{(m)},y)=I_m(x^{(m)},y,0)$ and since $I_m(x^{(m)},y,0)$ is lower semi-continuous in $y$
we obtain that $J_m(x^{(m)},y)$ is also lower semi-continuous in $y$, completing the induction step
in this case, as well.

 Now we can construct $\pi^*\in\Pi_x$ and $\sig^*\in\cT_{0N}$ so that (\ref{optimal}) holds true with
 such $\pi^*$ and $\sig^*$. Set $X_0^{\pi^*}=x$ and inductively
 \[
 X^{\pi^*}_{n+1}=X^{\pi^*}_n+\la_n(\rho^{(n)},X_n^{\pi^*})\Del S_{n+1},
 \]
 where
 \[
 \la_n(x^{(n)},y)=\mbox{argmin}_{\gam\in G_n(x^{(n)},y)}I_n(x^{(n)},y,\gam)
 \]
 if $a_{n+1}(x^{(n)})\ne 0$ and $b_{n+1}(x^{(n)})\ne 0$ while
 \[
 \la_n(x^{(n)},y)=I_n(x^{(n)},y,0)
 \]
 if $a_{n+1}(x^{(n)})=0$ or $b_{n+1}(x^{(n)})=0$, recalling that in this case
 $I_n(x^{(n)},y,z)$ does not depend on $z$. Thus, we set $\gam_{n+1}=\la(\rho^{(n)},X_n^{\pi^*})$
 and $\be_{n+1}=X_{n+1}^{\pi^*}-\gam_{n+1}S_{n+1}$. Define also
 \[
 \sig^*=\min\{ 0\leq n\leq N:\,\ell(U_n-X_n^{\pi^*})=\Psi_n^{\pi^*}\}.
 \]
 Verifying (\ref{optimal}) we prove by the backward induction that
 \begin{equation}\label{L10.4}
 J_n(\rho^{(n)},X_n^{\pi^*})=\Psi_n^{\pi^*}
 \end{equation}
 and that for any $\pi\in\Pi_x$,
 \begin{equation}\label{L10.5}
 J_n(\rho^{(n)},X_n^{\pi})\leq \Psi_n^{\pi}.
 \end{equation}

 By the definition $J_N(\rho^{(N)},X_N^\pi)=\Psi_N^\pi$ for any admissible self-financing strategy $\pi$, and so
 (\ref{L10.4}) and (\ref{L10.5}) are trivially satisfied for $n=N$. Suppose that (\ref{L10.4}) and (\ref{L10.5})
  hold true for all $N\geq n\geq m+1$ and prove them for $n=m$. Relying on the properties of regular conditional
   probabilities discussed above we can write that with probability one,
\begin{eqnarray*}
& E\big( J_{m+1}(\rho^{(m+1)},X_{m+1}^\pi)|\cF_m\big)=E\big( J_{m+1}((\rho^{(m)},\rho_{m+1}),\, X^\pi_{m}\\
&+\gam_{m+1}\rho_{m+1}\ka_m(\rho^{(m)})|\cF_m\big)
=\int_{-1}^\infty J_{m+1}((\rho^{(m)},u),\, X^\pi_{m}+\gam_{m+1}u\ka_m(\rho^{(m)}))d\mu_m(u).
 \end{eqnarray*}
It follows from the definition of $\Psi^\pi_m$, $I_m(\rho^{(m)},X_m^\pi,\gam_{m+1})$ and
$J_m(\rho^{(m)},X_m^\pi)$ that with probability one,
\[
\Psi_m^\pi\geq I_m(\rho^{(m)},X_m^\pi,\gam_{m+1})\geq J_m(\rho^{(m)},X_m^\pi)
\]
for any admissible self-financing strategy $\pi$, completing the induction step for (\ref{L10.5}).

 On the other hand, if we choose $\gam^*_{m+1}=\gam_{m+1}=\la(\rho^{(m)},X_m^{\pi^*})\in G(\rho^{(m)},
 X_m^{\pi^*})$ then by the construction of $\pi^*$,
 \[
 I_m(\rho^{(m)},X_m^{\pi^*},\gam^*_{m+1})=J_m(\rho^{(m)},X_m^{\pi^*}).
 \]
 By the induction hypothesis with probability one,
 \begin{eqnarray*}
 &I_m(\rho^{(m)},X_m^{\pi^*},\gam^*_{m+1})=\min\big(\ell(f_m(\rho^{(m)})-X_m^{\pi^*}),\max(\ell(g_m(\rho^{(m)})
 -X_m^{\pi^*}),\\
 & E(J_{m+1}(\rho^{(m+1)},X_{m+1}^{\pi^*})|\cF_m))\big)=\min\big(\ell(f_m(\rho^{(m)})-X_m^\pi),\\
 &\max(\ell(g_m(\rho^{(m)})-X_m^\pi),\, E(\Psi_{m+1}^{\pi^*}|\cF_m))\big)=\Psi_m^{\pi^*}.
  \end{eqnarray*}
 Hence, $J_m(\rho^{(m)},X_m^{\pi^*})=\Psi_m^{\pi^*}$ completing the induction step for (\ref{L10.4}).

 Observe that $\pi^*=(\be^*_n,\gam_n^*)_{n=1}^N$ with $\be^*_n=X_n^{\pi^*}-\gam_n^*S_n$. The formula for
 the optimal stopping time $\sig^*$ follows from the standard results about Dynkin's games which were
 discussed in Lecture 3. Finally,
 \[
 r(\pi^*,\sig^*)=r(\sig^*)=\Psi_0^{\pi^*}=J_0(S_0,x)\leq\Psi_0^\pi=r(\pi)
 \]
 for any admissible self-financing portfolio strategy $\pi\in\Pi_x$, and so (\ref{optimal}) holds true,
 completing the proof of the theorem.
 \end{proof}

\bibliography{matz_nonarticles,matz_articles}
\bibliographystyle{alpha}

\end{document}